\documentclass[11pt]{article}
\usepackage{moriond,epsfig}

\def\Journal#1#2#3#4{{#1} {\bf #2}, #3 (#4)}

\def\NPB{{\em Nucl. Phys.} B}
\def\PLB{{\em Phys. Lett.}  B}

\def\PRD{{\em Phys. Rev.} D}
\def\PR{\em Phys. Rev.}
\def\ZPC{{\em Z. Phys.} C}
\def\ANN{\em Ann. Phys.}
\def\RMP{\em Rev. Mod. Phys.}
\def\MPLA{{\em Mod. Phys. Lett.} A}
\def\PTP{\em Progr. Theor. Phys.}
\def\EPJ{{\em Eur. Phys. J.} C}

\def\be{\begin{equation}}
\def\ee{\end{equation}}
\def\bea{\begin{eqnarray}}
\def\eea{\end{eqnarray}}

\begin{document}
\vspace*{2cm}
\title{Coulomb and hadronic scattering in
elastic high-energy nucleon collisions}

\author{{\underline{V. Kundr\'{a}t}}, M. Lokaj\'{i}\v{c}ek }

\address{Institute of Physics AS CR, 182 25 Prague 8, Czech Republic}

\maketitle\abstracts{The commonly used West and Yennie model approach
to the description of the interference between Coulomb and hadronic scattering
of nucleons is critically examined and its deficiencies are clarified. The
preference of the more general eikonal model approach is summarized.}

High-energy elastic scattering of nucleons is realized 
mostly due to the strong hadronic interactions, and in the 
case of charged hadrons also by the Coulomb interactions
which are being commonly described with the help of the 
total elastic amplitude $F^{C+N}(s,t)$. This amplitude is
usually written as the sum of hadronic amplitude $F^{N}(s,t)$ 
and of Coulomb amplitude $F^{C}(s,t)$ being mutually 
correlated by a relative phase $\alpha \Phi (s,t)$ \cite{beth}: 
\begin{equation}
F^{C+N}(s,t) =
F^{C}(s,t) + F^{N}(s,t) e^{i\alpha \Phi(s,t)}, 
\label{wy1}
\end{equation}
where $\alpha=1/137.036$ is the fine structure constant; 
here $s$ is the square of CMS energy. For the phase
function $\Phi (s,t)$ West and Yennie \cite{west} have 
derived the formula 
\begin{equation}
\Phi (s,t) = \mp \bigg [ \ln \bigg ( {{-t} \over s} \bigg ) -
\int_{-4 p^2}^{0} {{ d t'} \over {|t - t'|}} \bigg ( 
1 - {{F^N(s,t')} \over {F^N(s,t)}} \bigg ) \bigg ],
\label{wy2}
\end{equation}
which contains an integral over all 
admissible  $t$  values and, therefore, has not 
been considered as an efficient tool for analysis 
of experimental data. In order to simplify it
the following assumptions for hadronic amplitude 
$F^N(s,t)$ have been accepted \cite {west}:
(i) the spins of all the particles involved 
can be neglected, (ii) the $t$ dependence of 
the modulus $|F^N(s,t)|$ is purely exponential 
in the whole kinematically allowed region of $t$
and (iii) both the real and imaginary parts of the 
$F^{N}(s,t)$ have the same $t$ dependence for
all admissible $t$ values.

Then the total elastic scattering amplitude $F^{C+N}(s,t)$
takes the form
\begin{equation}
F^{C+N}(s,t) =
\pm {\alpha s \over t} f_1(t)f_2(t)e^{i\alpha \Phi}+
{\sigma_{tot} \over {4\pi}} p\sqrt {s} (\rho+i)e^{Bt/2},
\label{wy4}
\end{equation}
where the phase $\alpha \Phi(s,t)$ equals 
\cite{west} 
\begin{equation}
\alpha \Phi (s,t) = \mp \alpha \bigg [ \ln \bigg ({{-B t} \over 2} 
\bigg ) + \gamma \bigg ].
\label{wy5}
\end{equation}
Here $\gamma = 0.577215$ is Euler's constant,
$p$ is the value of the momentum in CMS
and $\rho= {{Re F^N(s,t=0)}\over {\Im F^N(s,t=0)}}$.
Together with the diffraction slope $B$ the $\rho$
is assumed to be independent of $t$. 
Both these quantities together with the total cross 
section $\sigma_{tot}$ may be energy dependent only 
and may characterize the elastic hadron scattering at given 
energy. The two form factors $f_1(t)$ and $f_2(t)$ describe 
the space structure of colliding nucleons.  
The upper (lower) sign corresponds to the scattering 
of particles of the same (opposite) charges. 

It has been shown recently \cite{kun1} that
among the mentioned three assumptions the last one
is automatically involved in the requirement for
the relative phase $\alpha \Phi (s,t)$ given
by integral formula (\ref{wy2}) to be real. The 
reality of the phase requires immediately for the quantity
$\rho(s,t) \equiv {{\Re F^N(s,t)} \over {\Im F^N(s,t)}}$
to be constant for all admissible $t$ values. 

All other assumptions having played the role in 
the derivation of Eqs. (\ref{wy4}) and (\ref{wy5})
have been studied in detail \cite{kun1,kun2}. First, 
it has been shown that the existence of 
diffractive minimum is in a contradiction with 
the constant value of the quantity $\rho$. Second,
it has been pointed out that the exponential
$t$ dependence of the modulus of hadronic amplitude, 
i.e., $|F^N(s,t)| \sim e^{Bt/2}$ can be considered 
as being approximately satisfied only in the region
of $t$ running from zero to the position of diffractive 
minimum in the differential cross section. However, 
this position moves to $t=0$ when the energy increases. 
Thus the region of exponential behavior 
of the modulus $|F^N(s,t)|$ becomes narrower with
increasing energy. The deviations of the modulus 
$|F^N(s,t)|$ from the exponential behavior 
can be exhibited by the $t$ dependence of the 
diffractive slope $B(s,t)$ defined as
\begin{equation}
B(s,t)= {d\over {dt}}\bigg[\ln {d \sigma^{N}\over {dt}}\bigg] =
{2\over |F^{N}(s,t)|}{d\over {dt}}|F^{N}(s,t)|.
\label{sl1}
\end{equation}
It is evident that in the case of West and Yennie 
approach this quantity should be $t$ independent while
the experimental data exhibit its $t$ dependence \cite{kun3}. 
Thus, also the second assumption cannot be fulfilled at all 
kinematically allowed $t$ values as required in derivation 
of Eq. (\ref{wy5}). And we must conclude that both integral 
and simplified West and Yennie formulas contradict 
the elastic nucleon scattering differential cross 
section data.

The preference should be given to the eikonal model 
approach that is not burdened by similar limitations.
The elastic scattering amplitude can be defined
in this approach with the help of Fourier-Bessel 
transformation as
\begin{equation}
F(s,q^2=-t)= {s\over {4 \pi i}} \int\limits_{\Omega_b}d^2b
e^{i\vec{q}\vec{b}} \bigg[e^{2i\delta(s,b)}-1\bigg],
\label{ei1} 
\end{equation}
where $\delta(s,b)$ stands for the eikonal 
and $\Omega_b$ represents the two-dimensional 
Euclidean space of the impact parameter $\vec b$.
Mathematically consistent formulation of Fourier-Bessel 
transformation is guaranteed when the function 
$F(s,t)$ in the region of unphysical $t$ values 
is defined as analytical continuation from
the region of physical $t$ values in agreement with
formula (\ref{ei1}), comp. Adachi et al. \cite{adac} 
and Islam \cite{isla} who showed that it is then 
valid at any $s$ and $t$. 

The influence of both the Coulomb and strong interactions
can be described with the help of the sum of Coulomb and 
hadronic eikonals \cite{fran} and the  total elastic scattering 
amplitude may be written as
\begin{equation}
F^{C+N}(s,t=-q^2) = {s\over{4\pi i}} \int\limits_{\Omega_b} d^2b
e^{i\vec{q} \vec{b}}\bigg[e^{2i(\delta^{C}(s,b)
+\delta^{N}(s,b))}-1\bigg].
\label{ei2}
\end{equation}
Eq. (\ref{ei2}) can be then transformed   
\cite{fran} into the form
\begin{equation}
\!\!\!\!\!\!\!\! F^{C+N}(s,t) =   F^{C}(s,t) + F^{N}(s,t) +
{i\over {\pi s}} \int\limits_{\Omega_{q'}} d^2q'
F^{C}(s,{q'^2})F^{N}(s,[\vec{q} - \vec{q'}]^2),
\label{ei3}
\end{equation}
where $F^{C}(s,t)$ and $F^{N}(s,t)$ are Coulomb and elastic 
hadronic amplitudes defined by expression (\ref{ei1}) with
the eikonals $\delta^{C}(s,b)$ and $\delta^{N}(s,b)$. Eq. 
(\ref{ei3}) includes the convolution integral of the two 
amplitudes defined over kinematically allowed region 
of momentum transfers $\Omega_{q'}$.

Eq. (\ref{ei3}) can be rewriten as \cite{kun4}
\begin{equation}
F^{C+N}(s,t) = \pm {\alpha s\over t}f_1(t)f_2(t) +
F^{N}(s,t)\Bigg [1\mp i\alpha G(s,t) \Bigg ],  
\label{kl1}
\end{equation}
where
\begin{equation}
\!\!\!\!\!\!\!G(s,t) = \int\limits_{t_{min}}^0
dt'\Bigg \{ \ln \bigg( {t'\over t} \bigg )
{d \over{dt'}}
\bigg[f_1(t')f_2(t')\bigg]
+ {1\over {2\pi}}\Bigg [{F^{N}(s,t')\over F^{N}(s,t)}-1\Bigg]
I(t,t')\Bigg \},
\label{kl2}
\end{equation}
and
\begin{equation}
I(t,t')=\int\limits_0^{2\pi}d{\Phi^{\prime \prime}}
{f_1(t^{\prime \prime})f_2(t^{\prime \prime})\over t^{\prime \prime}};
\label{kl3}
\end{equation}
here $t^{\prime \prime}=t+t'+2\sqrt{tt'}\cos{\Phi}^{\prime \prime}$.
For the case of nucleon - nucleon scattering $t_{min}=-s + 4 m^2$.
Formulas (\ref{kl1}) - (\ref{kl3}) hold generally for any $s$ 
and $t$ up to the terms linear in $\alpha$.
The expression in the last bracket of Eq. (\ref{kl1})
may be regarded as the first term in the Taylor series
expansion of the exponential $e^{\mp i \alpha G}$; then
one can write within the same precision
\begin{equation}
F^{N+C}(s,t) = F^{C}(s,t) + F^{N}(s,t) e^{\mp i \alpha G(s,t)},
\label{to2}
\end{equation}
the form being analog of original 
formula (\ref{wy1}) of West and Yennie.
The $G(s,t)$ (being complex) cannot
be interpreted as a mere relative phase. 
The reality of $G(s,t)$ would be equivalent 
to require for the quantity $\rho(s,t)$ to be
constant and vice versa. 

Formulas (\ref{kl1}) - (\ref{kl3})  
can be used to determine the elastic hadronic amplitude 
from the differential cross section data provided its modulus 
and the phase are conveniently parameterized (for detail see
analysis of $pp$ and $p\bar{p}$ scattering at ISR and SPS 
energies \cite{kun4}) and for accurate prediction of the 
differential cross section at any $t$ if the hadronic amplitude
is given.

The analysis \cite{kun4} of the mentioned experimental data 
showed that: (i) the quantites $B(s,t)$ and $\rho(s,t)$ are $t$ 
dependent in the whole $t$ region,
(ii) the influence of Coulomb scattering cannot be neglected at 
higher $|t|$ values, either, and (iii) 
the peripheral picture of elastic nucleon scattering
seems to be slightly statistically preferred by experimental data.

The acurate determination of the elastic amplitude is also
important when the luminosity $\mathcal{L}$ should be calibrated
on the basis of elastic process; it holds \cite{bloc} 
\begin{equation}
{1 \over \mathcal{L}} {d N_{el} \over {dt}} = { \pi \over {s p^2}}
 |F^{C+N} (s,t)|^2,
\label{lu1}
\end{equation}
where ${d N_{el} \over {dt}}$ is the counting rate
established at corresponding $t$. The so called
Coulomb calibration in the region of smallest $|t|$
where the Coulomb amplitude is dominant (reaching nearly
100 $\%$) can be hardly realized at the LHC due to
technical problems. The approach allowing to avoid
corresponding difficulties may be based on Eq. (\ref{lu1}),
when the total elastic scattering amplitude $F^{C+N}(s,t)$ 
with required accuracy at the smallest achievable $|t|$ 
is established.

However, then it is very important which formula for the total
elastic amplitude $F^{C+N}(s,t)$ is made use of. In the
following we will demonstrate possible difference at different
$t$ values. We have calculated the quantity
\begin{equation}
R(t) =  {{|F^{C+N}_{eik}(s,t)|^2 -
|F^{C+N}_{WY}(s,t)|^2}
\over {|F^{C+N}_{eik}(s,t)|^2}} * 100, 
\label{lu2}
\end{equation}
where $F^{C+N}_{eik}(s,t)$ and $F^{C+N}_{WY}(s,t)$ are the
values of total elastic scattering amplitudes calculated 
in the framework of Bourelly-Soffer-Wu model \cite{bsw,tdr}
for $pp$ elastic scattering at the LHC energy 14 TeV
(the central behavior in the impact parameter space);
the former quantity with the help of Eqs. (\ref{kl1}) - (\ref{kl3}) 
and the latter one with the help of Eqs. (\ref{wy2}) - (\ref{wy4}). 
The $t$ dependence of the quantity $R(t)$ is represented in 
Fig. 1. Its maximum value is 0.8 $\%$ at $t=-0.006$ GeV$^2$ 
showing that the resulting differential cross sections 
determined with the help of commonly used West and Yennie 
method and with the help of mathematically and physically 
consistent eikonal approach differ. This difference is a function 
of the $t$ variable. Similar behavior of the quantity $R(t)$
has been also obtained in the preliminary analysis \cite{kasp} 
in the case of model of Islam et al.\cite{isl3} where
the maximum value of the quantity $R(t)$ is much higher.

$R(t)$ will also differ for the other
types of the phase of elastic hadronic amplitude - see 
the analysis of $pp$ and $p\bar{p}$ scattering at ISR and 
SPS energies \cite{kun4} corresponding to the central and 
peripheral distributions of elastic hadron scattering 
where this difference is yet larger in the peripheral 
case. It means that the luminosity determined for the 
central as well as peripheral distributions of elastic 
$pp$ scattering at LHC energy of 14 TeV may be burdoned 
by a non-negligible systematic error.
\begin{figure}
\epsfig{file=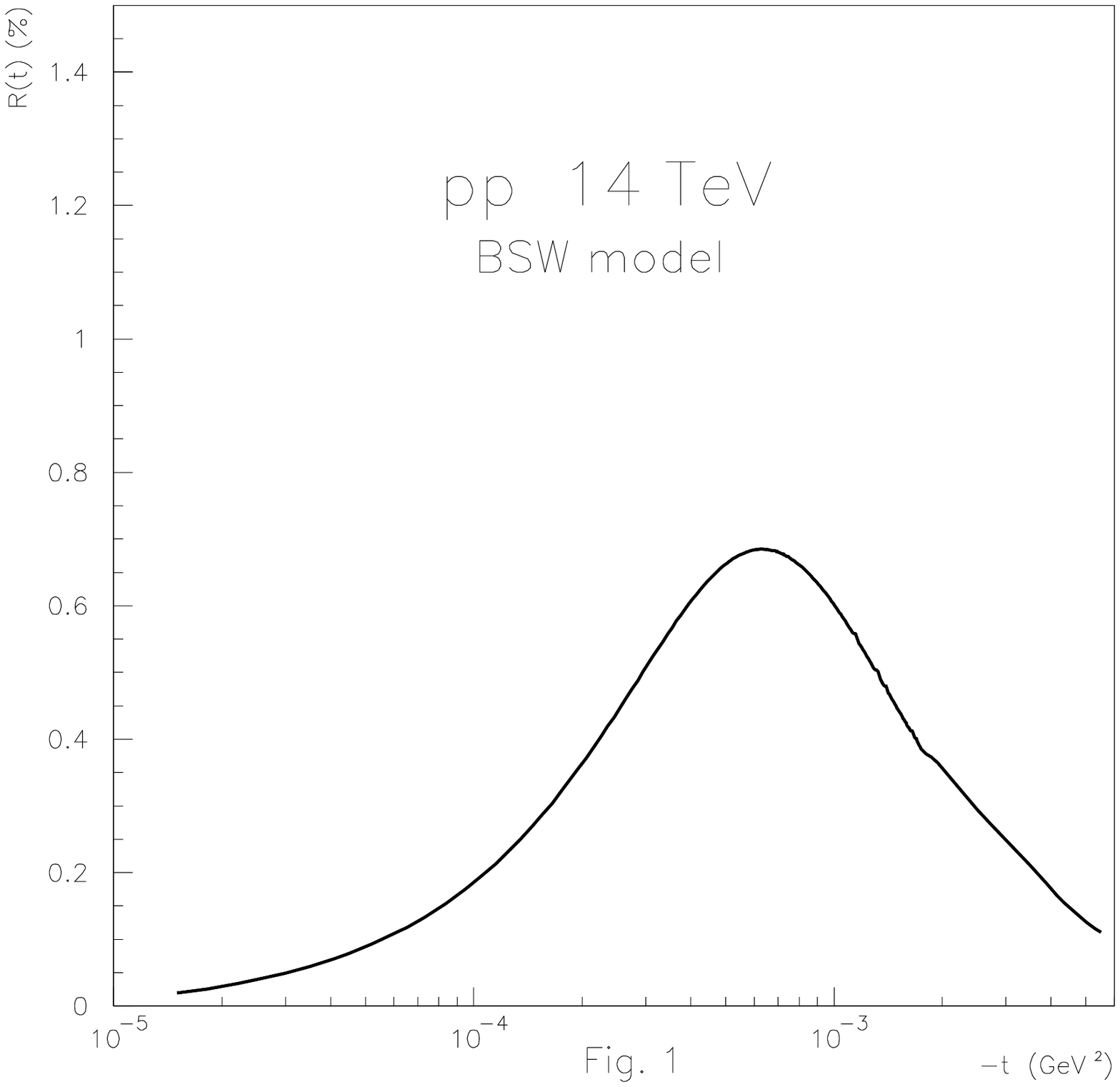,height=201pt,angle=0,width=230pt}
\vspace*{-2cm}
\caption{$t$ dependence of the quantity $R(t)$ defined by Eq.(\ref{lu2}) 
and calculated for the Bourrely-Soffer-Wu model} 
\caption{$t$ dependence of the same quantity $R(t)$ defined by Eq. (\ref{lu2}) 
and calculated for the model of Islam et al.}
\end{figure}

\end{document}